\documentclass[reprint, superscriptaddress,
 amsmath,amssymb,
 aps,
prl,
]{revtex4-2}

\usepackage{graphicx}
\usepackage{dcolumn}
\usepackage{bm}
\usepackage[colorlinks=true,allcolors=blue]{hyperref}

\begin{document}

\title{Transverse Spin Supercurrent at $p$-wave magnetic Josephson Junctions}

\author{Morteza Salehi}
 \affiliation{Physics Department, Bu-Ali Sina University, Hamadan, Iran}
 \email{m.salehi@basu.ac.ir}

\date{\today}

\begin{abstract}
We theoretically study a Josephson junction consisting of $s$-wave superconductors and a $p$-wave magnet. We find that in the presence of a strength vector of $p$-wave magnet, the electrons' and holes' dispersion relation shifts in the $k$-space. Additionally, we demonstrate that the perpendicular component of the strength vector converts Andreev bound states into Andreev modes that can propagate along the junction's interfaces. These modes create a transverse spin supercurrent while their transverse charge supercurrent is zero. These features open an opportunity to design superconducting spintronics devices.
\end{abstract}

\maketitle


\textit{Introduction} - The interplay between magnetism and superconductivity has led to the discovery of remarkable phenomena including spin-triplet superconductivity, topological superconducting states, and non-reciprocal charge transport \cite{Buzdin_RMP_2005, Bergeret2005RMP,volkov2003PRL, Mazin2007PRL,Sato_RPP_2017}. Recent advances in unconventional magnet ($u$M) has revealed new possibilities through materials like $d$-wave altermagnet ($d$AM) and $p$-wave magnet ($p$M) that combine zero net magnetization with non-relativistic spin splitting \cite{Smejkal_SciAdv_2022,Mazin2022PRX,Jungwirth2024ArXiv,Jungwirth2024ArXiv2,Smejkal_PRX_2022,Smejkal2022PRX,Hellens2024ArXiv,Song2025Nature,Linder2024PRL}.

The $u$Ms are characterized by zero net magnetization and their parity-dependent spin splitting: $d$-wave types preserve inversion symmetry while breaking time-reversal symmetry, whereas $p$-wave types maintain time-reversal symmetry but break inversion symmetry \cite{Hellens2024ArXiv,Sivianes2025PRL}. Although magnets can be used to produce spin current, their magnetization bring challenges for spin transport\cite{Zutic2004RMP}. In contrast, the $u$Ms are promising candidates for generating spin currents\cite{Chen2025AM}. In superconducting junctions, the $d$AMs \cite{Brekke2024PRB,Zhang2024NC} have been shown to induce 0-$\pi$ transitions \cite{Lu2024PRL,Linder2023PRL,Beenakker2023PRB,Sun2025PRB,Cheng2024PRB}, diode effect\cite{Sim2024ArXiv,Banerjee2024PRB,Cheng2024PRB2}, oriented-dependent Andreev reflection\cite{Linder2023PRB,Papaj2023PRB,Nagae2025PRB}, crossed Andreev reflection\cite{Soori2024PRB,Niu2024SST}, and generate odd-frequency pairing \cite{Chakraborty2024ArXiv}. Also, the interplay of  $p$Ms and superconductors are attracting attentions\cite{Fukaya2025PRB,Maeda2025PRB,Maeda2024JPSJ,Nagae2025ArXiv,Hong2025PRB,Sun2025Arxiv}.

In this Letter, we theoretically study a Josephson junction formed by s-wave superconductors and a $p$M. The $p$M strength vector splits spin-dependent bands, and its transverse component generates propagating Andreev modes along the junction interfaces. These modes carry a transverse spin supercurrent without a charge counterpart, offering a platform for dissipationless spin transport in superconducting spintronics\cite{Linder2015NP}.

\begin{figure}
\includegraphics*[scale=0.25]{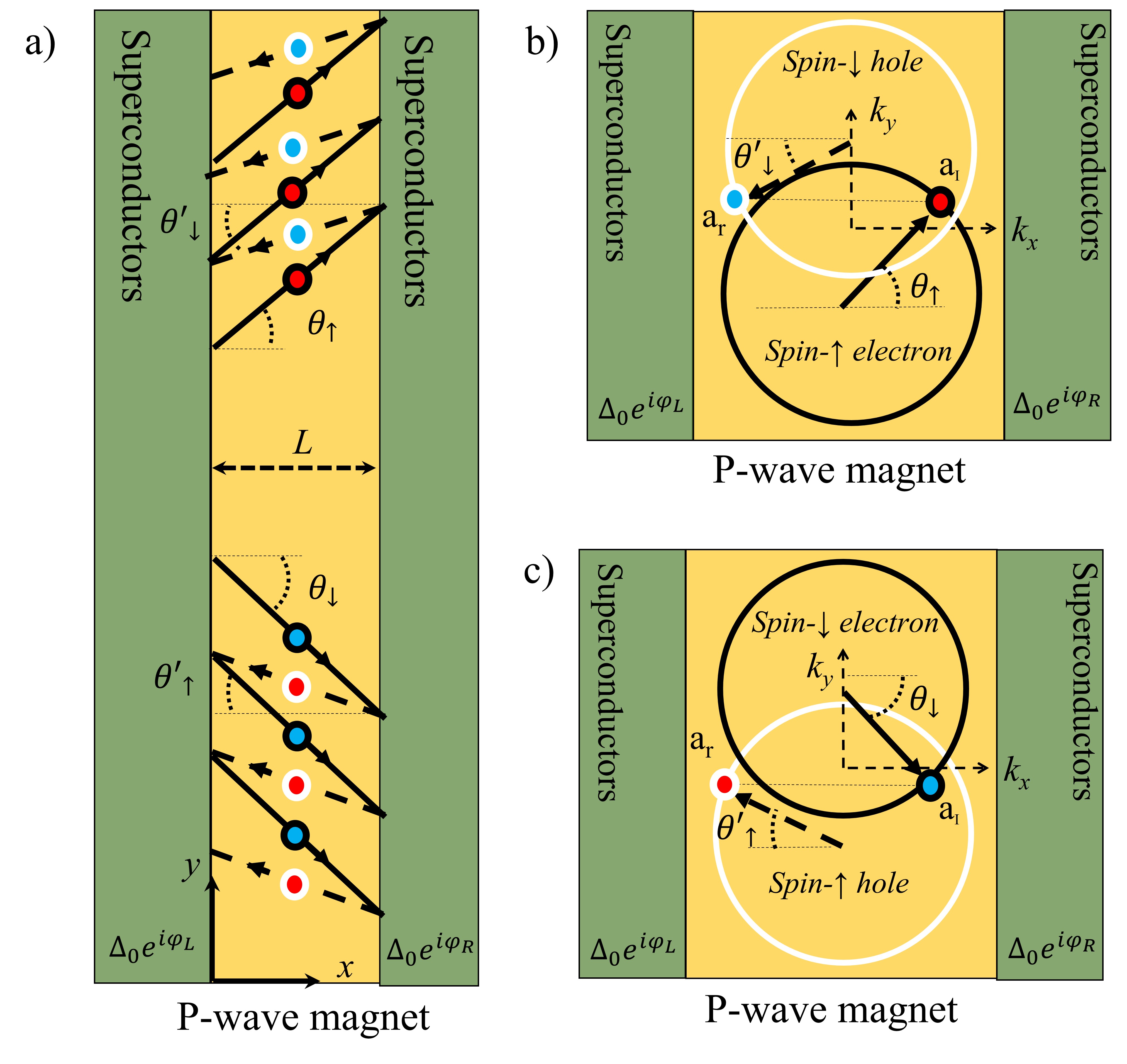}
\caption{(a) The p-wave magnet Josephson junction in real space. The black solid arrows show the propagation direction of electron particles, where the black dashed arrows demonstrate the reflected hole particles in the $p$M region. The black circles with red (blue) dots demonstrate the electron-like quasi particles with spin-$\uparrow (\downarrow)$ configuration. The white circles with red (blue) dots illustrate hole particles with spin-$\uparrow (\downarrow)$ configuration. The $\alpha_y$ creates the Andreev modes that propagate parallel to the interface of the junction. (b) The spin-$\uparrow$ electrons' and spin-$\downarrow$ holes' Fermi circles in the $k$-space that shift related to each other in the presence of $\boldsymbol{\alpha}\neq 0$. These Andreev modes produce a positive spin supercurrent in the $y$-direction of real space. Here, we set $(\alpha_x=0, \alpha_y \neq 0)$. (c) The spin-$\downarrow$ electrons' and spin-$\uparrow$ holes' Fermi circles that produce negative spin supercurrent in the $-y$-direction of real space.}
\label{Fig.k-dispersion}
\end{figure}

\emph{Model and formalism--- }As shown in Fig. \ref{Fig.k-dispersion}(a), our 2D model consists of a $p$M between two semi-infinite s-wave superconductors. We employ the Bogoliubov-de Gennes (BdG) formalism to obtain the Andreev bound states and their related charge and spin supercurrents\cite{deGennes1999}. In the absence of spin-mixing potential, the s-wave superconducting pair potential, $\Delta_0$, couples the spin-$\uparrow$ electrons with spin-$\downarrow$ holes and vice versa. So, the BdG Hamiltonian is \cite{Maeda2024JPSJ,deGennes1999,Sukhachov2025PRB,Maeda2025PRB,Nagae2025ArXiv},

\begin{equation}
 \mathcal{H}_{\pm}(\boldsymbol{k})=\left(
 \begin{array}{cc}
    \frac{\hbar^2}{2m}(\boldsymbol{k}\pm\boldsymbol{\alpha})^2-\mu  & \Delta_0 e^{i\phi}\\
  \Delta_0 e^{-i\phi}    & \mu- \frac{\hbar^2}{2m}(\boldsymbol{k}\mp\boldsymbol{\alpha})^2
 \end{array}\right).   
 \label{Eq.HBdG}
\end{equation}
The $\mathcal{H}_+(\textbf{k})$ acts on basis $\boldsymbol{\Psi}=(\psi_\uparrow, -\psi_\downarrow^*)$ while the $\mathcal{H_-(\textbf{k})}$ works with $\boldsymbol{\Psi}=(\psi_\downarrow, \psi_\uparrow^*)$. Also, the 2D wave vector, $\textbf{k}=(k_x, k_y)$, hosts the $\boldsymbol{\alpha}$ as $p$M strength vector. The $\mu$ is a Fermi energy that can be different in S and $p$M regions. Since $\phi$ stands for superconductivity phase, we define $\delta\phi=\phi_R-\phi_L$ as superconducting phase difference between right and left superconductors that produces supercurrents\cite{Josephson1962PL}. We take the junction in $x$-direction where $p$M region, $(\boldsymbol{\alpha}\neq 0 , \Delta_0 =0)$, is located at $0 \leq x \leq L$. The eigenvalues of $\mathcal{H}_+(\boldsymbol{k})$ for spin-$\uparrow$ electron excitations in the $p$M region are,
\begin{equation}
	\epsilon_e=(\hbar^2 \left((k_x+\alpha_x)^2+(k_y+\alpha_y)^2\right)/2m)-\mu.
	\label{Eq.eigenvalues1}
\end{equation}.
This dispersion relation shows a parabola in the $k$-space. For a fixed energy, the states of spin-$\uparrow$ electrons form a Fermi circle with the radius of $q_e=\sqrt{2m(\epsilon_e+\mu)/\hbar^2}$, centered at $(-\alpha_x, -\alpha_y)$, as depicted schematically in Fig.\ref{Fig.k-dispersion} (b). The propagation direction of each state in real space can be obtained via $\textbf{V}_e=\partial \epsilon_{e}/\hbar\partial\textbf{k}$. The state $a_I$ located on the spin-$\uparrow$ electron's Fermi circle moves with the angle of $\theta_{\uparrow} $ with respect to the $x$-direction in real space. The spin-$\downarrow$ holes of $\mathcal{H}_{+} (\boldsymbol{k})$ have energy dispersion of $\epsilon_h=(-\hbar^2 \left((k'_x-\alpha_x)^2+(k'_y-\alpha_y)^2\right)/2m)+\mu$. This relation gives rise to another Fermi circle with radius $q_h = \sqrt{2m|\epsilon - \mu|/\hbar^2}$, centered at $(\alpha_x, \alpha_y)$ for a fixed energy value. In the same scenario, their propagation angle, $\theta'_\downarrow$, are determined by the group velocity, $\textbf{V}_h=-\partial \epsilon_h/\hbar\partial\textbf{k}$. 

The spin-$\uparrow$ hitting electron to the superconductor interface from the non-superconducting side of the Josephson junction can be reflected as a spin-$\downarrow$ hole during the Andreev process to transfer an s-wave Cooper pair into the superconducting lead\cite{Andreev1964}. The reflected hole can be back-scattered again as an electron from the other surface of the Josephson junction and creates an Andreev-bound state in non-superconducting region \cite{deGennes1964PL,Deutscher2005RMP}. In the ballistic limit, the energy and parallel component of the wave vector with respect to the junction's interface, $k_y$, are conserved during the scattering processes. In the presence of $\alpha_y\neq 0$, where the electrons' and holes' Fermi circles shift in $k_y$-direction oppositely, the Andreev process occurs just in the overlap zone of two Fermi circles\cite{Salehi2023PS}. As shown in Fig.\ref{Fig.k-dispersion}(a), the Andreev bound states convert to Andreev modes and propagate parallel to the junction's interfaces. \cite{Titov2007PRB,Salehi2025JMMM}. The contribution of each Andreev mode to the transverse charge supercurrent is proportional to $\sim (\partial \varepsilon_+/\partial\delta\phi)(\sin\theta_{\uparrow}+\mathcal{S}\sin\theta'_{\downarrow})$. Also, its share to the transverse spin supercurrent is $\sim (\partial \varepsilon_+/\partial\delta\phi)(\sin\theta_{\uparrow}-\mathcal{S}\sin\theta'_{\downarrow})$. Here, $\mathcal{S}=sign(\mu-\epsilon)$ determines the location of the reflected hole in conduction or valence band\cite{Beenakker2006PRL}. Also, $\varepsilon_+(\delta\phi)$ is the energy-phase relation (EPR) of Andreev modes.

On the other hand, $\mathcal{H}_-(\boldsymbol{k})$ has a similar scenario with spin$\downarrow$ electron and its related spin-$\uparrow$ hole. In contrast to spin-$\uparrow$ electrons, the dispersion relation of spin-$\downarrow$ electrons is centered at $(\alpha_x, \alpha_y)$. As depicted in Fig.\ref{Fig.k-dispersion}(c),  the origin of the spin-$\uparrow$ holes' dispersion relation is at $(-\alpha_x, -\alpha_y)$. Due to this reverse adjustment of spin-dependent dispersion relation in $k$-space, the Andreev modes with EPR of $\varepsilon_-(\delta\phi)$, propagate opposite to the Andreev modes with EPR of $\varepsilon_{+}(\delta\phi)$. The transverse charge supercurrent of $\varepsilon_-(\delta\phi)$ cancels the share of $\varepsilon_+(\delta\phi)$, whereas the resulting transverse spin supercurrent of $\varepsilon_-(\delta\phi)$ sums up the share of $\varepsilon_+(\delta\phi)$ with a factor of 2. As shown in Fig.\ref{Fig.k-dispersion}(a), a pure transverse spin supercurrent flows parallel to the junction interfaces, whereas there is no transverse charge supercurrent.

The overlap of two Fermi circles determines the critical angles of $\theta_{\uparrow}$, where within those values the Andreev modes are real and stable. Since the parallel component of the wave vector, $k'_y=k_y$, and the energy of excitations, $\epsilon_e=\epsilon_h=\epsilon$ are conserved during the Andreev process, we use the $x$-component of the hole wave vector to obtain the critical angles in the $\alpha_y\neq 0$ case,
\begin{equation}
    k'_x=\pm\sqrt{\frac{2m}{\hbar^2}(\mu-\epsilon)-(k_y-\alpha_y)^2}.
    \label{Eq.k_hx}
\end{equation}
The $k'_x$ must be real to have stable Andreev modes. From $k_y=q_e \sin\theta_{\uparrow}-\alpha_y$, the lower and upper critical angles of Andreev modes for $\mathcal{H}_+(\boldsymbol{k})$ can be obtained. In a similar way for $\mathcal{H}_-(\boldsymbol{k})$ we have,
\begin{equation}
    \begin{array}{l}
    \theta^\pm_{min}=\sin^{-1}\left(\text{max}\{-1,\frac{-q_h\pm 2\alpha_y}{q_e}\}\right)  \\
     \theta^\pm_{max}=\sin^{-1}\left(\text{min}\{1,\frac{q_h\pm 2\alpha_y}{q_e}\}\right).
    \end{array}
    \label{Eq.ThetaCritical}
\end{equation}
To obtain $\varepsilon_+(\delta\phi)$ from $\mathcal{H}_+(\textbf{k})$, we begin with the right and left mover wave functions of spin-$\uparrow$ electrons in $p$M region,
\begin{equation}
    \psi_{e,\uparrow}^{\pm}(\textbf{r})=\frac{1}{\sqrt{V_{e,x}}}\left(
    \begin{array}{l}
         1  \\
         0 
    \end{array}
    \right)e^{\pm i q_e \cos\theta_{\uparrow}x-i\alpha_x x}e^{i k_yy}.
    \label{Eq.Psi_e}
\end{equation}
Here, $V_{e,x}$ is the $x$-component of the group velocity and acts as a normalization factor of probability conservation. The $\pm$ sign refers to the propagation direction with respect to $ x$-axis in real space. Also, the spin-$\downarrow$ hole wave functions are,
\begin{equation}
    \psi_{h,\downarrow}^{\pm}(\textbf{r})=\frac{1}{\sqrt{V_{h,x}}}\left( \begin{array}{l}
         0  \\
         1 
    \end{array}\right)e^{\mp i q_h \cos\theta'_{\downarrow}x+i \alpha_x x}e^{i k_y y}.
\end{equation}
Due to the different scattering processes, the whole wave function in the $p$M region is\cite{Maiti2007PRB},
\begin{equation}
    \Psi_{p\text{M}}(\textbf{r})=a_1 \psi_{e,\uparrow}^+(\textbf{r})+a_2 \psi_{e,\uparrow}^-(\textbf{r})+a_3 \psi_{h,\downarrow}^+(\textbf{r})+a_4 \psi_{h,\downarrow}^+(\textbf{r}).
\end{equation}
One can calculate the electron-like and hole-like excitations of the superconducting region, $(\boldsymbol{\alpha=0}, \Delta_0\neq 0)$, as below,
\begin{equation}
    \psi_e^{S,\pm}(\textbf{r})=\frac{1}{\sqrt{V_s}}\left(\begin{array}{l}
         e^{i\beta}  \\
         e^{-i\phi}
    \end{array}
    \right)e^{\pm i k^S_{e,x}x}e^{i k_yy}
    \nonumber
\end{equation}
\begin{equation}
    \psi_h^{S,\pm}(\textbf{r})=\frac{1}{\sqrt{V_s}}\left(\begin{array}{l}
         e^{i\phi}  \\
         e^{i\beta}
    \end{array}
    \right)e^{\mp i k^S_{h,x}x}e^{i k_yy}.
    \label{Eq.Psi_S}
\end{equation}
We define $\beta=\cos^{-1}(\epsilon/\Delta_0)$. Also, $k^S_{e (h),x}$ and $V_S$ are the $x$-component of electron (hole) wave vector and group velocity in the superconductor region, respectively. In a similar way, we writes the whole wave functions of right and left superconductor,
\begin{equation}
\begin{array}{l}
	 \Psi^S_L(\textbf{r})=a_5 \psi^{S,-}_e(\textbf{r})+a_6\psi^{S,-}_h(\textbf{r}) \\
     \Psi^S_R(\textbf{r})=a_7 \psi^{S,+}_e(\textbf{r})+a_8\psi^{S,+}_h(\textbf{r})  
    
\end{array}    .
\label{Eq.Psi_S}
\end{equation}
Here, $a_j$ with $j=\{1, ... , 8\}$ is the related probability amplitude of each moving wave function.
The first boundary condition confirms the continuity of wave functions at interfaces,
\begin{equation}
    \begin{array}{l}
       \Psi^S_L(x=0)=\Psi_{p\text{M}}(x=0) \\
        \Psi_{p\text{M}}(x=L)=\Psi^S_R(x=L),
    \end{array},
    \label{Eq.BC1}
\end{equation}
while the second boundary condition satisfies the conservation of probability distribution\cite{Maeda2024JPSJ,Sepkhanov2007,Salehi2015AP},
\begin{equation}
     \begin{array}{l}
       V^S \Psi^S_L(x=0)=V_{e (h),x}\Psi_{p\text{M}}(x=0)  \\
       V_{e (h),x}\Psi_{p\text{M}}(x=L)=V^S \Psi^S_R(x=0)
    \end{array}.
    \label{Eq.BC2}
\end{equation}

\begin{figure}
\includegraphics*[scale=0.4]{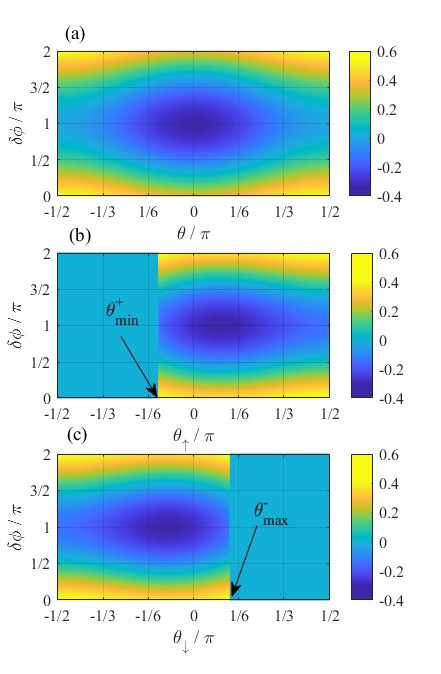}
\caption{(a) The EPR of the Josephson junction in the absence of $p$M strength vector. In this case, the $\varepsilon_{+}(\delta\phi)$ and $\varepsilon_-(\delta\phi)$ are degenerate. Here, we set $\mu=\Delta_0$, $\mu_S=10 \Delta_0$ and $L=\kappa_0^{-1}$. (b) The $\varepsilon_{+}(\delta\phi) $ with respect to the propagation direction of spin-$\uparrow$ incoming electron. The EPR becomes imaginary for $\theta_\uparrow < \theta^+_{min}$. (c) The $\varepsilon_-(\delta\phi)$ with respect to the propagation angle of spin-$\downarrow$ incoming electrons. Here, $\varepsilon_-(\delta\phi)$ becomes evanescent for $\theta^-_{max} < \theta_\downarrow$. Also, we set $\alpha_y = 0.3  \kappa_0$.}
\label{Fig.Andreev modes}
\end{figure}
\textit{EPR--} The energy-phase relation of the Josephson junction can be obtained via the Eqs (\ref{Eq.BC1}) and (\ref{Eq.BC2}). They lead to eight homogeneous equations for probability amplitudes of $a_j$. To have nontrivial solutions, the determinant of its coefficient matrix must be zero. We focus on the short-junction regime $(\epsilon \ll \mu)$ to have experimentally possible results. The transverse spin supercurrent emerges in the presence of $\alpha_y \neq 0$. With these assumptions, the EPR of $\mathcal{H}_+(\boldsymbol{k})$ can be calculated with the simplification of the determinant such as

	\begin{equation}
			\varepsilon_{+}(\delta\phi)=\Delta_0 \cos\left(
		\begin{array}{l}
			\frac{1}{2}\cos^{-1}\left(\frac{\mathcal{A}\cos\delta\phi-\mathcal{D}}{\sqrt{\mathcal{B}^2+\mathcal{C}^2}}\right)\\
			+\frac{1}{2}\cos^{-1}\left(\frac{\mathcal{B}}{\sqrt{\mathcal{B}^2+\mathcal{C}^2}}\right)
			\end{array}\right).
			\label{Eq.EPR1}
	\end{equation}

where we use,

\begin{equation}
	\begin{array}{l}
		\mathcal{A} =4q^2 \cos\theta_{\uparrow}\cos\theta'_{\downarrow}, \\
		\\
		\begin{array}{ll}
			\mathcal{B}= & 4 q^2\cos(k_{x}L)\cos (k'_{x}L)\cos\theta_{\uparrow}\cos\theta'_{\downarrow} \\
			& +(1+q^2\cos^2\theta_{\uparrow})(1+q^2\cos^2\theta'_{\downarrow})\sin(k_{x}L)\sin(k'_{x}L),\\
			&
		\end{array}\\
	\begin{array}{ll}
		\mathcal{C}= & 2q (1+q^2\cos^2\theta_{\uparrow})\cos\theta'_{\downarrow}\sin(k_{x}L)\cos(k'_{x}L)\\
		&+ 2q (1+q^2\cos\theta'_{\downarrow})\cos\theta_{\uparrow}\cos(k_{x}L)\sin(k'_{x}L),
	\end{array} \\
\\
	\begin{array}{ll}
		\mathcal{D}=&- (1-q^2\cos^2\theta_{\uparrow})(1-q^2\cos^2\theta'_{\downarrow})\sin(k_{x}L)\sin(k'_{x}L).\\
	
	\end{array}
	\end{array}
\end{equation}
Here, $q=\sqrt{\mu/\mu_s}$ is the square root of the relative Fermi energies. Also, $k_{x}$ can be obtain from Eq.(\ref{Eq.eigenvalues1}). As shown in part (a) of Fig.(\ref{Fig.k-dispersion}), the $\varepsilon_{+}(\delta\phi)$ and $\varepsilon_-(\delta\phi)$ are degenerate in the absence of the $p$M strength vector. Since $\alpha_x$ does not contribute to the transverse spin supercurrent, we set $\alpha_x=0$. We normalize all energy scales with the superconducting gap, $\Delta_0$. Moreover, the superconducting wave vector, $\kappa_0=(2m\Delta_0/\hbar^2)^{1/2}$, is used to normalize the wave vectors. In the presence of $\alpha_y\neq 0$, the electrons' and holes' Fermi circles shift oppositely in the $k_y$-direction and overlaps partially. The Andreev modes are created in this overlap zone, and their propagation directions are confined between the thresholds defined by Eq.(\ref{Eq.ThetaCritical}). In part (b) and (c) of Fig.(\ref{Fig.Andreev modes}), the EPRs of $\mathcal{H}_+(\boldsymbol{k})$ and $\mathcal{H}_-(\boldsymbol{k})$ are depicted. For spin-$\uparrow$ electrons with propagation direction beyond the critical values of Eq.(\ref{Eq.ThetaCritical}), the $\varepsilon_{+}(\delta\phi)$ becomes evanescent and cannot participate in the transverse charge and spin supercurrents.  Similarly, $\varepsilon_-(\delta\phi)$ cannot propagate in $\theta^-_{max} < \theta_\downarrow$. 

\begin{figure}
\includegraphics*[scale=0.35]{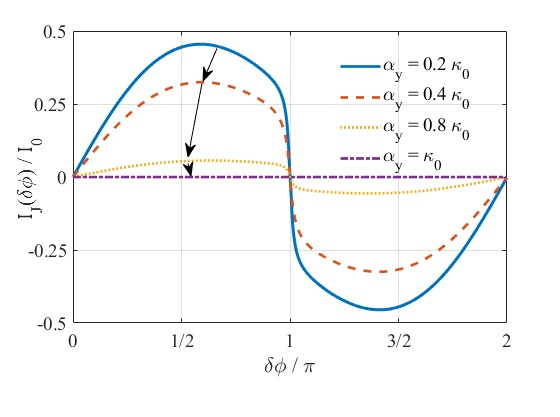}
\caption{The charge supercurrent that passes through the Josephson junction vs superconducting phase difference. An increase in $\alpha_y$ decreases the charge supercurrent. Further increase, $\kappa_0 \leq \alpha_y$, does not lead to $0-\pi$ transition. The other inputs are same as Fig.(\ref{Fig.Andreev modes}.)}
\label{Fig.ChargeSuperCurrent}
\end{figure}

\textit{Charge Supercurrent--} The charge supercurrent at zero temperature that passes through the junction interfaces is given by \cite{Titov2007PRB,Salehi2025JMMM,Linder2008PRL}
\begin{equation}
	\frac{I_J(\delta\phi)}{I_0}=\sum_{\gamma=\pm}\int_{-\frac{\pi}{2}}^{\frac{\pi}{2}}\frac{-\partial \varepsilon_\gamma(\delta\phi)}{\partial \delta\phi}(\cos\theta_{\uparrow(\downarrow)}+\cos\theta'_{\downarrow(\uparrow)})d\theta_{\uparrow(\downarrow)}.
	\label{Eq.LChargeSuperCurrent}
\end{equation}
Here, $I_0=2e\Delta_0\mathcal{N}(\mu)/\hbar$ is the normalized ballistic supercurrent, where $\mathcal{N(\mu)}$ is the density of states at the Fermi energy. The Andreev modes of $\varepsilon_{\gamma}(\delta\phi)$ have two parts: incoming electrons with propagation direction of $\theta_{\uparrow(\downarrow)}$ and reflecting holes with propagation direction of $\theta'_{\downarrow(\uparrow)}$. The $\cos\theta_{\uparrow(\downarrow)}$ determines the share of the first one with respect to the $x$-direction, whereas the $\cos\theta'_{\downarrow(\uparrow)}$ takes the share of last part into account. In the short-junction regime, the Andreev modes have a dominant role in the Josephson current. For incoming electrons with propagation direction beyond the critical angles of Eq.(\ref{Eq.ThetaCritical}), there is a negligible chance of co-tunneling for Cooper pairs to pass the Josephson junction that we ignore in our calculations\cite{Titov2007PRB,Furusaki1991SSC}. The charge supercurrent that flows in the junction is depicted in Fig.(\ref{Fig.ChargeSuperCurrent}) for different values of $\alpha_y$. An increase in $\alpha_y$ reduces the overlapping region between the electron and hole Fermi circles, as depicted in Fig.(\ref{Fig.k-dispersion}). This also reduces the charge supercurrent that flows in the junction. For $ k_F \leq \alpha_y$, where $k_F=(2m \mu/\hbar^2)^{1/2}$, the overlap zone disappears and the charge supercurrent tends to zero. More increase in $\alpha_y$ cannot change the sign of charge supercurrent and we do not encounter the $\pi$-junction\cite{Buzdin1999PRB}.  The transverse charge supercurrent can be calculated via,
\begin{equation}
		\frac{I^t_J(\delta\phi)}{I_0}= \sum_{\gamma=\pm}\int_{\theta^\gamma_{min}}^{\theta^\gamma_{max}}\frac{-\partial \varepsilon_\gamma(\delta\phi)}{\partial \delta\phi}(\sin\theta_{\uparrow(\downarrow)}+\mathcal{S}\sin\theta'_{\downarrow(\uparrow)})d\theta_{\uparrow(\downarrow)}.
	\label{Eq.TChargeSuperCurrent}
\end{equation}
Here, the contribution of electron and hole parts of Andreev modes with respect to the $y$-direction is calculated by $\sin\theta_{\uparrow(\downarrow)}$ and $\sin\theta'_{\downarrow(\uparrow)}$, respectively. As its obvious from Fig.(\ref{Fig.k-dispersion}) and can be calculated from Eq.(\ref{Eq.TChargeSuperCurrent}), the transverse charge supercurrent of $\varepsilon_{+}(\delta\phi)$ cancel out with the contribution of $\varepsilon_{-}(\delta\phi)$. It means there is no transverse charge supercurrent flowing in the $p$M region of the Josephson junction.

\begin{figure}
	\includegraphics[scale=0.35]{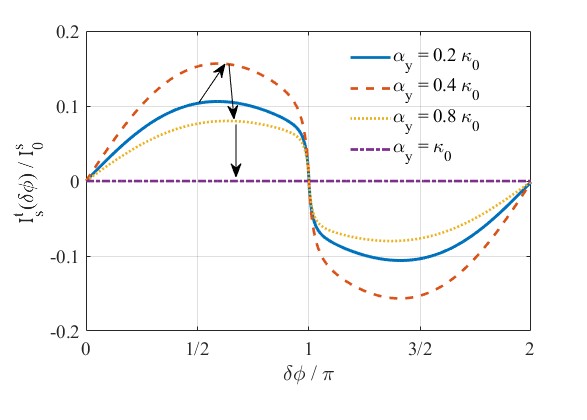}
	\caption{The transverse spin supercurrent that flows parallel to the Josephson junction's interfaces. Appearance of $\alpha_y$ creates a transverse spin supercurrent. The normalized spin supercurrent is $I^s_0=\hbar I_0/2$. The other inputs are same as Fig.(\ref{Fig.Andreev modes}.)}
	\label{Fig.TSSvsphase}
\end{figure}

\textit{Spin Supercurrent--} The electronic part of each Andreev mode carry the amount of $ (-\partial\varepsilon_{\gamma}/\partial \delta\phi)\gamma \hbar\cos\theta_{\uparrow(\downarrow)}/2$ spin supercurrent in $x$-direction whereas this quantity for hole part is $ (-\partial\varepsilon_{\gamma}/\partial \delta\phi)\bar{\gamma} \hbar\cos\theta'_{\downarrow(\uparrow)}/2$. So, the spin supercurrent in the $x$-direction can be calculated via,
 
 \begin{equation}
 		\frac{I_s(\delta\phi)}{I^s_0}=\sum_{\gamma=\pm}\int_{-\frac{\pi}{2}}^{\frac{\pi}{2}}\gamma\frac{-\partial \varepsilon_\gamma(\delta\phi)}{\partial \delta\phi}(\cos\theta_{\uparrow(\downarrow)}-\cos\theta_{\downarrow(\uparrow)})d\theta_{\uparrow(\downarrow)}.
 	\label{Eq.LSpinSuperCurrent}
 \end{equation}
Here, the normalized ballistic spin supercurrent is $I^s_0=\hbar I_0/2e$. Due to the s-wave character of two superconducting leads, there is no spin supercurrent to flow in the $x$-direction of the junction.

On the other hand, the contribution of $\varepsilon_{+}(\delta\phi)$ to spin supercurrent in $y$-direction is given by, $ \hbar(-\partial\varepsilon_{+}/\partial \delta\phi)( \sin\theta_{\uparrow}-\mathcal{S}\sin\theta'\downarrow)/2$ while the contribution that comes from $\varepsilon_{-}(\delta\phi)$ is equal $ \hbar(-\partial\varepsilon_{-}/\partial \delta\phi)( \sin\theta_{\downarrow}-\mathcal{S}\sin\theta'_\uparrow)/2$. So, the transverse spin supercurrent in $y$-direction is given by,

\begin{equation}
	 		\frac{I^t_s(\delta\phi)}{I^s_0}=\sum_{\gamma=\pm}\int_{\theta^\gamma_{min}}^{\theta^\gamma_{max}}\gamma\frac{-\partial \varepsilon_\gamma(\delta\phi)}{\partial \delta\phi}(\sin\theta_{\uparrow(\downarrow)}-\mathcal{S}\sin\theta_{\downarrow(\uparrow)})d\theta_{\uparrow(\downarrow)}.
	\label{Eq.TSpinSuperCurrent}
\end{equation}

\begin{figure}
	\includegraphics[scale=0.35]{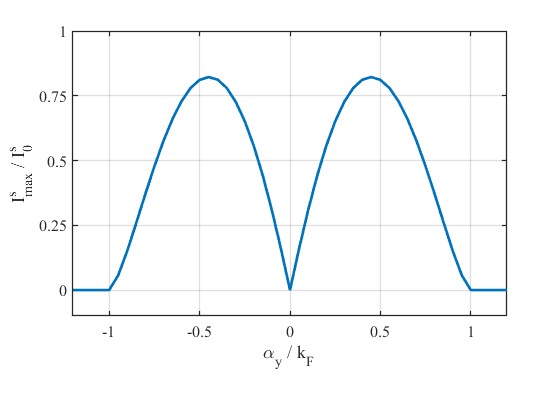}
	\caption{The critical values of transverse spin supercurrent vs $\alpha_y$.}
	\label{Fig.TSSvsAlpha}
\end{figure}

In the presence of $ 0 \leq \alpha_y$, only the Andreev modes of $\varepsilon_{+}(\delta\phi)$ survive in the range of $\theta^-_{max}\leq \theta_\uparrow \leq \theta^+_{max}$ while the Andreev modes of $\varepsilon_{-}(\delta\phi)$ is evanescent. This creates a positive spin supercurrent in $y$-direction. A reverse situation occurs in the range of $\theta^+_{min}\leq \theta_\downarrow \leq \theta^-_{min}$ where the Andreev modes of $\varepsilon_{+}(\delta\phi)$ become evanescent while their $\varepsilon_{-}(\delta\phi)$ counterparts are propagating. This leads to a negative spin supercurrent in $-y$-direction. So, the summation of these two terms with the zero value of Eq.(\ref{Eq.TChargeSuperCurrent}) for transverse charge supercurrent leads to a purely transverse spin supercurrent that flows in the $p$M region parallel with Josephson's interfaces. We illustrate the transverse spin supercurrent for different values of $\alpha_y$ in Fig.(\ref{Fig.TSSvsphase}). As $\alpha_y$ increases from zero, the transverse spin supercurrent in the junction begins to flow. With a further increase in $\alpha_y$, the spin supercurrent increases and reaches its maximum. As shown in Fig.(\ref{Fig.TSSvsAlpha}), beyond this point, $\alpha_y \sim k_F /2$, further increase in $\alpha_y$ leads to a decrease in the spin supercurrent.

Now, we aim to predict the results that can be obtained in experiments. The typical length of a ballistic Josephson junction is $ L \sim 1\,\mu\text{m}$, and we assume the width to be $ W \sim 1\,\mu\text{m} $. The typical value of the superconducting gap is $ \Delta_0 \sim 1\,\text{meV}$, and the Fermi energy is assumed to be comparable to the superconducting gap. This leads to a Fermi wave vector of approximately $ k_F \sim 1.33 \times 10^8\,\text{m}^{-1}$. Accordingly, the available density of modes in the junction is estimated as $ \mathcal{N}(\mu) =k_F W /\pi \sim 50$. This implies that the ballistic charge supercurrent is approximately $ I_0 \sim 100\,\mu\text{A}$. Based on Fig.(\ref{Fig.ChargeSuperCurrent}), the predicted charge supercurrent flowing through the $p$M Josephson junction is in the range $ I_J \sim 10\text{--}50\,\mu\text{A} $. Finally, the estimated transverse spin supercurrent flowing parallel to the Josephson interfaces is \( I_s^t \sim 3.1 \times 10^6\,\hbar/\text{s} \), which is fully detectable using currently available instruments.

\textit{Conclusion}-- To summarize, we have studied the unconventional $p$-wave magnetic Josephson junction. It is found that electrons' and holes' dispersion relations shift in $k$-space in the presence of $p$-wave magnet strength vector. The perpendicular component of $\boldsymbol{\alpha}$ converts the Andreev bound states into Andreev modes that can propagate alongside the Junction's interfaces. We show that these modes create a transverse spin supercurrent while the transverse charge supercurrent is zero. We believe this prediction can be detected with available technology and can be used to design future superconducting spintronics devices.

\textit{Acknowledgment}-- M. S. thanks R. Beiranvand for fruitful discussion.

\end{document}